\documentclass[preprint,showpacs,showkeys,preprintnumbers,amsmath,amssymb]{revtex4}

\usepackage{graphicx}
\usepackage{dcolumn}
\usepackage{bm}

\begin{document}

\author{R. Rossi Jr.}
\affiliation{Universidade Federal de S\~{a}o Jo\~{a}o del-Rei,
Campus Alto Paraopeba, C.P. 131, 36420–000, Ouro Branco, MG, Brazil}

\title{Entanglement preservation on two coupled cavities}

\begin{abstract}
The dynamics of two coupled modes sharing one excitation is
considered. A scheme to inhibit the evolution of any initial state
in subspace $\{|1_{a},0_{b}\rangle , |0_{a},1_{b}\rangle\}$ is
presented. The scheme is based on the unitary interactions with an
auxiliary subsystem, and it can be used to preserve the initial
entanglement of the system.
\end{abstract}

\maketitle

Idealized scenarios where one can manipulate individual atoms or
photons are essential ingredients for the development of Quantum
Theory. These thought-experiments were consider in the early days as
simple abstractions, very useful for theoretical proposes but could
never be implemented in real laboratories. However, considerable
technological development on the field of cavity QED, ions trap,
Josephson junction, have allowed for observations of interactions
between fragile quantum elements such as single photons, atoms and
electrons. Some examples are the interactions between Josephson
qubits \cite{art1}, single emitting quantum dot and radiation field
\cite{art2} and between Rydberg atoms and a single mode inside a
microwave cavity \cite{art3}. Such remarkable experimental control
opened the possibility for experimental investigations on
foundations of quantum mechanics. Recent examples are
\cite{art4,art5,art6}.

Beside this fundamental issues, the present technological scenario
provides means for possible revolutionary advances such as the
quantum computation, which is known to be extremely more powerful
than classical computation \cite{art7}. Inspired by this possible
revolutionary technological achievement several different strategies
were proposed to control, manipulate and protect quantum states.
Some examples are error-avoiding \cite{art8} and error-correcting
codes \cite{art9}, bang-bang control \cite{art10}, Super-Zeno effect
\cite{art11}. The well-known Quantum Zeno Effect, which was first
presented in the literature as a paradoxal consequence of
measurements on quantum mechanics \cite{art12}, is also a useful
tool for quantum state protection \cite{art13}, entanglement control
\cite{art14} and entanglement preservation \cite{art15}.

In Ref.\cite{art16} the Quantum Zeno effect in a bipartite system,
composed of two couple microwave cavities ($A$ and $B$), is studied.
It is shown how to inhibit the transition of a single photon,
prepared initially in cavity $A$, by measuring the number of photons
on cavity $B$. The measurement of the photon number is performed by
a sequence of $N$ resonant interactions between the cavity $B$ and
two level Rydberg atoms. As $N\rightarrow\infty$ the transition
inhibition became complete, and the initial state $|1,0\rangle$ is
preserved. However, an entangled state as
$a|1,0\rangle+b|0,1\rangle$ can not be preserved with such Quantum
Zeno scheme.

In the present work, it is shown a scheme to preserve any initial
state in subspace $\{|1_{a},0_{b}\rangle , |0_{a},1_{b}\rangle\}$.
The scheme is based on unitary interactions between the system of
interest and an auxiliary subsystem. An advantage of the present
scheme is that the procedure does not depend on the initial state.
It is also shown how to preserve the entanglement on subspace
$\{|1_{a},0_{b}\rangle , |0_{a},1_{b}\rangle\}$.

Let us consider the operators $\sigma_{x}, \sigma_{y}$ and
$\sigma_{z}$ in subspace $\{|1_{a},0_{b}\rangle ,
|0_{a},1_{b}\rangle\}$:

\begin{eqnarray}
\sigma_{x}=|1,0\rangle\langle 0,1| + |0,1\rangle\langle 1,0|,\\
\sigma_{y}=i\left(|1,0\rangle\langle 0,1| - |0,1\rangle\langle
1,0|\right),\\ \sigma_{z}=|1,0\rangle\langle 1,0|
-|0,1\rangle\langle 0,1|.
\end{eqnarray}

A general Hamiltonian in such subspace can be written as:

\begin{equation}
H=\frac{\hbar\omega}{2}\vec{S}\cdot\hat{n},\label{ham}
\end{equation}
where $\vec{S}\cdot\hat{n}=\left(\sigma_{x} \hat{i} +
\sigma_{y}\hat{j} + \sigma_{z}\hat{k}\right)\hat{n}$ is the spin
observable along the unit vector $\hat{n}=\sin\theta\cos\phi \hat{i}
+ \sin\theta\sin\phi \hat{j} + \cos\theta \hat{k}$, characterized by
the polar angles $\theta$ and $\phi$. The unitary operator,
$U_{n}(t)=e^{-\frac{iHt}{\hbar}}$, that represents the evolution
governed by the Hamiltonian (\ref{ham}) can be written, in the base
$\{|1_{a},0_{b}\rangle , |0_{a},1_{b}\rangle\}$, as

\begin{equation}
  U_{n}(t) = \left[
    \begin{array}{cc}
      [\cos(\omega t)+i\sin(\omega t)]-2i\cos^{2}\frac{\theta}{2}\sin(\omega t) &  2ie^{-i\phi}\cos\frac{\theta}{2}\sin\frac{\theta}{2}\sin(\omega t) \\
       2ie^{-i\phi}\cos\frac{\theta}{2}\sin\frac{\theta}{2}\sin(\omega t) & [\cos(\omega t)+i\sin(\omega t)]-2i\cos^{2}\frac{\theta}{2}\sin(\omega t)
    \end{array} \right].\label{Un}
\end{equation}

The fundamental aspect of the present quantum state control scheme
relays on the fact that when $\theta=k\frac{\pi}{2}$ (were $k$ is an
odd number) we can write:

\begin{equation}
\sigma_{z}U_{n,\frac{\pi}{2}}(t)\sigma_{z}=U_{n,\frac{\pi}{2}}(-t),\label{UUkU}
\end{equation}
where $U_{n,\frac{\pi}{2}}$ denotes the unitary evolution operator
(\ref{Un}) when $\theta=k\frac{\pi}{2}$. Therefore with a simple
procedure it is possible to construct an operator that can reverse
quantum state evolution. Using these operations we can control the
vector state dynamics restricting it to a certain trajectory on
Bloch sphere.

If a even number ($N$) of $\sigma_{z}$ operations are performed
periodically in a time interval $T$, the quantum state evolution
will be written as

\begin{equation}
|\psi(T)\rangle_{N}=\left[\sigma_{z}U_{n,\frac{\pi}{2}}\left(\frac{T}{N}\right)\right]^{N}|\psi(0)\rangle=\left[U_{n,\frac{\pi}{2}}\left(-\frac{T}{N}\right)U_{n,\frac{\pi}{2}}\left(\frac{T}{N}\right)\right]^{N/2}|\psi(0)\rangle=|\psi(0)\rangle.
\end{equation}

In the end the evolved quantum state is brought back to the initial
state. These sequence of operations can maintained the vector state
evolution in certain trajectory over the Bloch Sphere during the
time interval $T$. Notice that such procedure does not depend on the
initial state.

It is shown next that we can use this scheme to control an entangled
state dynamics and preserve the inicial concurrence. As the scheme
allows for the control of a quantum state in a two level system
subspace, we restrict the investigation for entangled states in
subspace $\{|1_{a},0_{b}\rangle , |0_{a},1_{b}\rangle\}$.

To make the analysis concrete let us consider the physical system
composed by two coupled modes ($M_{a}$ and $M_{b}$) sharing one
excitation. The Hamiltonian for the system is given by
\begin{equation}
H_{ab}=\hbar \omega a^{\dagger }a+\hbar \omega b^{\dagger }b+\hbar
g(a^{\dagger }b+b^{\dagger }a),  \label{Hamiltonian}
\end{equation}
where $a^{\dagger }$ ($a$) and $b^{\dagger }$ ($b$) are creation
(annihilation) operators for modes $M_{a}$ and $M_{b}$, $\omega $ is
their frequency and $g$ the coupling constant. As the modes share
only one excitation the dynamics is limited to subspace
$\{|1_{a},0_{b}\rangle , |0_{a},1_{b}\rangle\}$.

A implementation for such interaction can be realized in the context
of microwave cavity. Experimental proposals involving couple
microwave cavities are reported in Ref.\cite{art17,art18}. In Ref.
\cite{art17} the cavities are coupled by a conducting wire (wave
guide), and in Ref.\cite{art18} the cavities are connected by a
coupling hole. For both proposals the coupling allows the photon to
tunnel between the cavities, and the hamiltonian that governs such
dynamics is written in equation (\ref{Hamiltonian}).

The time evolution operator $U_{S}(t)=e^{\frac{-iH_{ab}}{\hbar}t}$
can be written in the base $\{|1_{a},0_{b}\rangle ,
|0_{a},1_{b}\rangle\}$ as:

\begin{equation}
  U_{S}(t) = \left[
    \begin{array}{cc}
      \cos(\omega t) &  -i\sin(\omega t) \\
       -i\sin(\omega t) & \cos(\omega t)
    \end{array} \right],\label{Un2}
\end{equation}
notice that the operator (\ref{Un2}) is equal to operator (\ref{Un})
with $\theta=k\frac{\pi}{2}$ (this is an essencial condition for the
control scheme) and $\phi=0$.

The initial state
$|\psi(0)\rangle=\cos\left(\frac{\theta_{0}}{2}\right)|1_{a},0_{b}\rangle+e^{i\phi_{0}}\sin\left(\frac{\theta_{0}}{2}\right)|0_{a},1_{b}\rangle$,
has the time evolution given by:

\begin{equation}
U_{S}(t)|\psi(0)\rangle=\alpha(t)|1,0\rangle+\beta(t)|0,1\rangle,
\end{equation}
where

\begin{eqnarray}
\alpha(t)&=&\cos\left(\frac{\theta_{0}}{2}\right)\cos gt
-ie^{i\phi_{0}}\sin\left(\frac{\theta_{0}}{2}\right)\sin gt,\\
\beta(t)&=& -i\cos\left(\frac{\theta_{0}}{2}\right)\sin gt
+e^{i\phi_{0}}\sin\left(\frac{\theta_{0}}{2}\right)\cos gt.
\end{eqnarray}

To study the entanglement dynamics between $M_{a}$ and $M_{b}$ the
concurrence $C(t)$, is calculated

\begin{equation}
C(t)=|\alpha^{\ast}(t)\beta(t)|
\end{equation}
for a detailed calculation of the concurrence see Ref.\cite{art19}.

It is possible to inhibit the evolution of the initial state and
consequently preserve the initial entanglement using the scheme
describe previously. It is clear that an essential ingredient for
such scheme is the sequence of $\sigma_{z}$ operations dividing the
unitary evolution. Let us now show that the interactions between the
present system and an auxiliary subsystem have the same effect as
the $\sigma_{z}$ operations.

For the physical system of two coupled cavities an adequate
auxiliary subsystem can be composed of a set of two level Rydberg
atoms (whose states are represented by $|e^{(k)}\rangle$ and
$|g^{(k)}\rangle$), that cross the cavity $B$, one at the time,
interacting with mode $M_{b}$ through a controlled time interval.
Each interaction is described by the Jaynes-Cummings model, and the
interaction hamiltonian can written as

\begin{equation}
H^{(k)}_{SA}=I_{a}\otimes \gamma \hbar
(b^{\dagger}|g^{(k)}\rangle\langle e^{(k)}| +
b|e^{(k)}\rangle\langle g^{(k)}|),
\end{equation}
where $\gamma$ is the coupling constant. A well known result of the
Jaynes-Cummings model is that when the interaction time is
$\tau=\frac{2\pi}{\gamma}$ we have

\begin{eqnarray}
U^{(k)}(\tau)|0_{b}\rangle|g^{k}\rangle&=&|0_{b}\rangle|g^{k}\rangle\\
U^{(k)}(\tau)|1_{b}\rangle|g^{k}\rangle&=&-|1_{b}\rangle|g^{k}\rangle,
\end{eqnarray}
where $U^{(k)}$ denotes the time evolution operator of the $k$-th
interaction between $M_{b}$ and the auxiliary subsystem. Therefore,
the time evolution governed by $U^{(k)}(\tau)$ act as $\sigma_{z}$
in subspace $\{|1_{a},0_{b}\rangle , |0_{a},1_{b}\rangle\}$ if the
atom is prepared in the ground state, as it is shown:

\begin{eqnarray}
U^{(k)}(\tau)\left(|1_{a}\rangle|0_{b}\rangle\right)|g^{k}\rangle&=&\left(|1_{a}\rangle|0_{b}\rangle\right)|g^{k}\rangle,\label{op1}\\
U^{(k)}(\tau)\left(|0_{a}\rangle|1_{b}\rangle\right)|g^{k}\rangle&=&-\left(|0_{a}\rangle|1_{b}\rangle\right)|g^{k}\rangle.\label{op2}
\end{eqnarray}

The time of interaction between the atoms and $M_{b}$ can be
controlled by stark effect, as in Ref. \cite{art20}. Therefore it is
possible to set the time of interaction between each atom and the
mode $M_{b}$ to be $\tau=\frac{2\pi}{g}$, which corresponds to a
$\pi$ pulse and preforms the operations (\ref{op1}) and (\ref{op2}).
For the Rubydium atoms used in the experiment \cite{art21}, the $\pi
$ Rabi pulse time is $\tau _{\pi }\simeq 10^{-5}s$. Let us consider
the time of interaction between $M_{a}$ and $M_{b}$ as
$T=\frac{\pi}{2g}$. In the experimental proposal of Ref.\cite{art17}
it was estimated the value for the coupling constant $g=10^{3}$,
therefore $T\simeq 10^{-3}s$. For simplicity let us consider each
interaction between $M_{b}$ and the two level atoms as
instantaneous, which is a good approximation as $\tau\ll T$ ( or
equivalently $g\ll \gamma$).

The sequence of operations $U^{(k)}(\tau)U_{S}(t)U(^{(k)}(\tau)$ has
the same effect of the operations in equation (\ref{UUkU}) on the
subspace $\{|1_{a},0_{b}\rangle , |0_{a},1_{b}\rangle\}$.

A control for the time evolution of the concurrence in time interval
$T$ can be performed if $T$ is divided by $N$ interactions between
$M_{b}$ and the auxiliary subsystem. This controlled time evolution
is composed of free evolutions of subsystem $M_{a}$-$M_{b}$,
governed by the unitary operator $U_{S}$, divided by $N$
instantaneous interactions with two level Rydberg atoms prepared in
the ground state, described by $U^{(k)}$. The time evolution can be
written as:

\begin{equation}
|\psi(T)\rangle_{N}=\left[U^{(N)}(\tau)U_{S}\left(\frac{T}{N}\right)\right]\ldots\left[U^{(k)}(\tau)U_{S}\left(\frac{T}{N}\right)\right]\left[U^{(k-1)}(\tau)U_{S}\left(\frac{T}{N}\right)\right]\ldots\left[U^{(1)}(\tau)U_{S}\left(\frac{T}{N}\right)\right]|\psi(0)\rangle.
\end{equation}

The total evolution is divided in $N$ steps, each one composed by a
free evolution $U_{S}$ and a interaction $U^{(k)}(\tau)$. After an
even number of interactions the vector state evolution can be
written as

\begin{equation}
\left[U^{(j)}(\tau)U_{S}\left(\frac{T}{N}\right)\right]\left[U^{(j-1)}(\tau)U_{S}\left(\frac{T}{N}\right)\right]\ldots\left[U^{(1)}(\tau)U_{S}\left(\frac{T}{N}\right)\right]|\psi(0)\rangle
=|\psi(0)\rangle\label{UUS}
\end{equation}
where $j$ is even. After an even number of interactions the state
vector is brought back to the initial state, as mentioned before,
therefore, the concurrence is given by $C(Tj/N)=C(0)$.

After an odd number of interactions ($j+1$), the state vector can be
written as:

\begin{equation}
\left[U^{(j+1)}(\tau)U_{S}\left(\frac{T}{N}\right)\right]\left[U^{(j)}(\tau)U_{S}\left(\frac{T}{N}\right)\right]\ldots\left(U^{(1)}(\tau)U_{S}\left(\frac{T}{N}\right)\right)|\psi(0)\rangle
=U^{(j+1)}(\tau)U_{S}\left(\frac{T}{N}\right)|\psi(0)\rangle,
\end{equation}

The concurrence of the system does not change with the operation
$U^{(k)}(\tau)$. Therefore, the concurrence after an odd number of
interactions is equal to the concurrence $C(T/N)$ of the state
$|\psi\left(\frac{T}{N}\right)\rangle=
U_{S}\left(\frac{T}{N}\right)|\psi(0)\rangle$.

The sequence of operations represented in equation (\ref{UUS}) can
be used to control the concurrence of the system $M_{a}$-$M_{b}$. In
the time $T$,  in which the sequence of operations is performed, the
concurrence is forced to oscillate between $C(0)$ (after an even
number of interactions) and $C\left(\frac{T}{N}\right)$(after an odd
number of interactions).

To ilustraste such effect let us consider the curve on Fig. 1, where
the concurrence of the system $M_{a}$-$M_{b}$ is represented as a
function of $t$. The initial state evolves freely and when $gt=0.3$
undergoes an interaction with the auxiliary subsystem. Notice that
for the initial state
$\frac{1}{\sqrt{2}}\left(|1,0\rangle+|0,1\rangle\right)$ the
concurrence decrease if no interactions with the auxiliary subsystem
is performed (see the thick line). However, if an interaction
$U^{(k)}(\tau)$ is performed, the concurrence starts to increase and
assumes the initial value when $gt=0.6$.

If $N$ interactions are performed, the control illustrated in Fig.1
for one interaction proceed and the concurrence is restricted to the
interval $C(0)\leq C\leq C(T/N)$. Notice that

\begin{equation}
\lim_{N\rightarrow \infty}C(T/N)=C(0),
\end{equation}
therefore, if the number of interactions increase in a finite time
interval $T$, the concurrence approaches to the constat value
$C(0)$, the initial concurrence, as it is shown in Fig.2.

To summarize, in the present work it is shown a scheme to control
the unitary dynamics of any initial state in the subspace
$\{|1_{a},0_{b}\rangle , |0_{a},1_{b}\rangle\}$. The scheme allows
for the inhibition of the concurrence evolution, preserving the
initial entanglement of the system.

\begin{figure}[h]
\centering
  \includegraphics[scale=0.5]{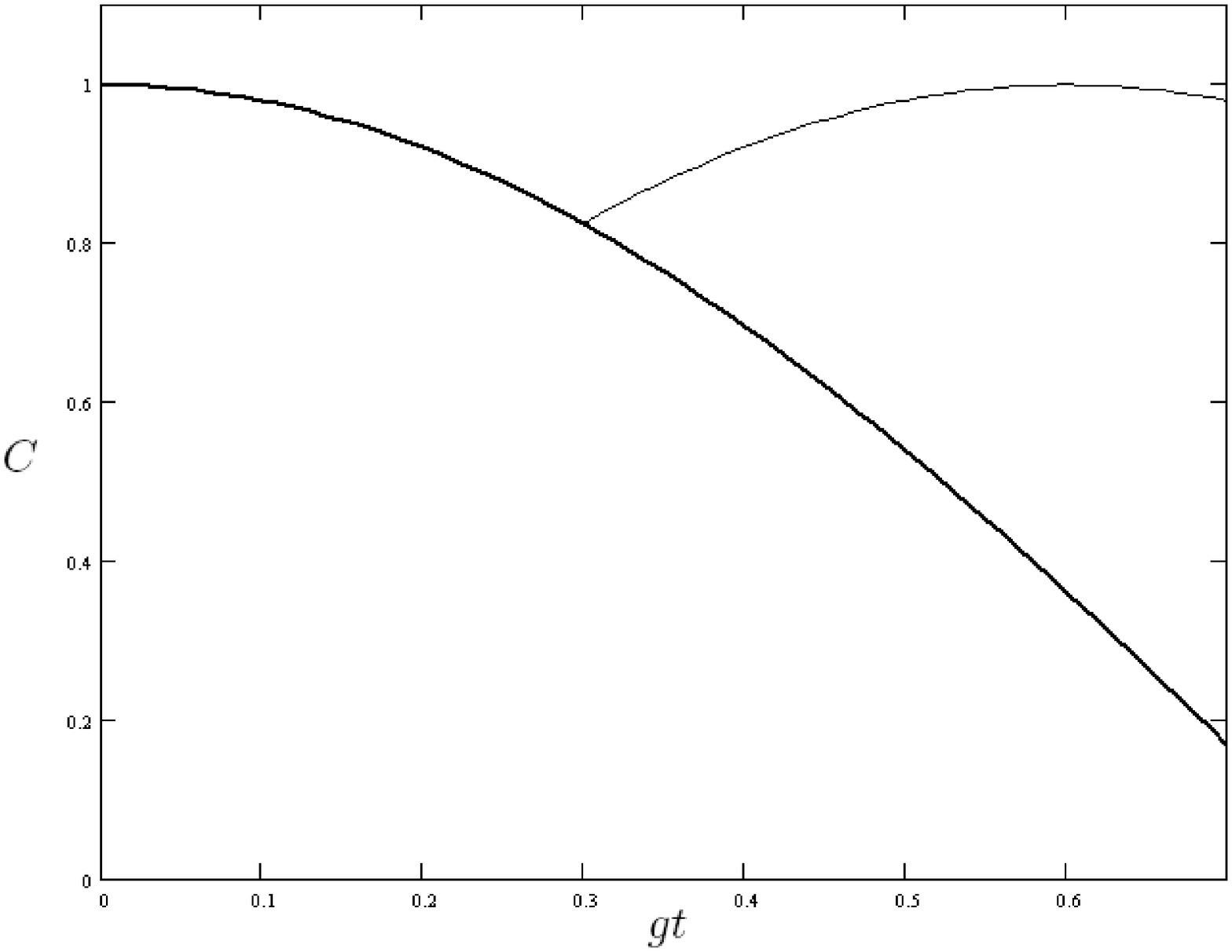}\\
  \caption{$C\times gt$, without interactions between $M_{b}$ and the atom (thick line) and with an interaction between one atom and $M_{b}$ at $gt=0.3$ (thin line).}
\end{figure}

\begin{figure}[h]
\centering
  \includegraphics[scale=0.5]{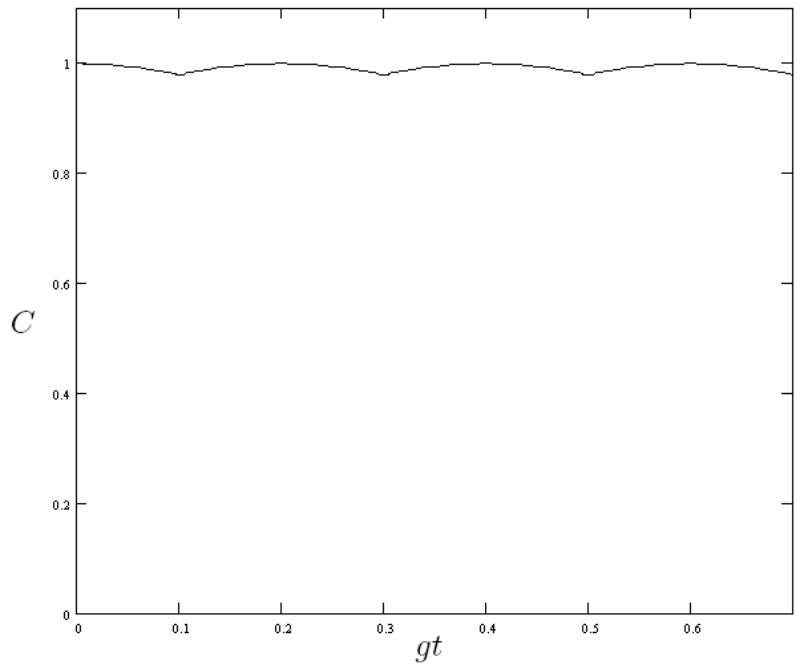}\\
  \caption{$C\times gt$, with interactions between $M_{b}$ and three atom at $gt=0.1$, $gt=0.2$ and $gt=0.3$ .}
\end{figure}

\end{document}